\pdfoutput=1
\documentclass[11pt]{article}
\usepackage[title]{appendix}
\usepackage{epsfig,latexsym}
\usepackage{bookmark}
\usepackage{longtablex}
\usepackage{amsmath}
\usepackage{longtable}
\usepackage{adjustbox}
\usepackage{hyperref}
\usepackage{verbatim}
\usepackage{color}
\usepackage{mathrsfs}
\usepackage{slashed}
\usepackage{needspace}
\usepackage{amssymb}
\usepackage{array}
\usepackage{jcappub}
\usepackage[utf8]{inputenc}

\usepackage{xcolor}
\usepackage{physics}
\usepackage{graphicx} 

\numberwithin{equation}{section}

\begin{document}

\title{On Filaments, Prolate Halos and Rotation Curves}

\author{K.~Zatrimaylov}
\emailAdd{kirill.zatrimaylov@sns.it}
\affiliation{\emph{Scuola Normale Superiore and INFN}\\\emph{Piazza dei Cavalieri 7, 56126, Pisa, Italy}}

\abstract{We propose a simple geometrical mechanism for the flattening of galactic rotation curves, the local compression of field lines around their planes induced either by the presence of thin string--like objects at the centers of galaxies or by elongated dark--matter halos, and elaborate on its possible role in Nature. We fit 83 rotation curves from the SPARC database with logarithmic potentials produced by a thin ``wire`` at the origin and then, after selecting 2 galaxies that yield the most interesting fits, analyze them with an alternative model, deformed versions of two popular models of dark--matter halos. Our conclusion is that the presence of a filament clearly improves the fit quality in a number of cases, while bulged dark matter profiles have a lesser effect. If taken at face value, these results would imply the presence of elongated mass distributions away from the galactic plane in a number of galaxies, and may also have some indirect impact on the controversy between cold dark matter (CDM), self--interacting dark matter (SiDM), and modified Newtonian dynamics (MOND).
}

\maketitle

\section{Introduction}\label{sec:introduction}
Historically, the detection of flattened galaxy rotation curves~\cite{Rubin:1970zza,Sofue:2000jx} was a key piece of evidence pointing to a dark--matter component in the Universe. The proposals to account for this phenomenon range from the so-called isothermal sphere DM distributions, which give rise to logarithmic gravitational potentials~\cite{BinneyTremaine}, to modifications of Newton's laws at small accelerations, such as MOND (modified Newtonian dynamics)~\cite{Milgrom:1983ca}.
However, in most cases the rotation curves are not quite flat. Rather, they tend to have either slightly decreasing or slightly increasing profiles at large distances~\cite{Persic:1995ru,Salucci:2007tm}, which suggests that their flattening is more likely the result of dynamical effects than of new physical laws.

In this paper we focus on a possible dynamical effect of this type, which apparently was not examined in detail before. Our message is that \emph{standard physical laws can suffice to yield flattened rotation curves in the presence of a long, thin filament at the galaxy center} via an effect whose origin would thus be purely geometric. An object of this type, observed at the center of Milky Way, has been hypothesized to be a jet of relativistic particles produced by the black hole~\cite{morris}. A more exotic candidate would be a cosmic string~\cite{morris}: in the models of the early Universe that include both primordial black holes (PBHs) and cosmic strings, strings can attach themselves to PBHs and migrate to the center of galaxies~\cite{Vilenkin:2018zol}. Cosmic strings have also been proposed as the cause for the distortions in pulsar signals observed by the NANOGrav collaboration; if true, this result may also lend indirect support to the conjecture of their presence in galaxies~\cite{Blasi:2020mfx,Arzoumanian:2020vkk}.

We also give some consideration to another effect that can contribute to the flattening of rotation curves at large distances, namely the prolate shape of the dark matter halo itself. Indeed, a number of galaxies and galaxy clusters, including Andromeda (M31) and the Milky Way, were estimated to have prolate dark halos, on the basis of kinematical data~\cite{Hayashi:2014nra,Bowden:2016bwq,Hattori:2019lgu} or gravitational lensing~\cite{Hoekstra:2003pn,Oguri:2004in}. If dark matter density distributions, which for simplicity we still assume rotationally symmetric in the planes of galaxies, were elongated in the orthogonal directions, the resulting quasi-logarithmic potentials would yield near-constant rotation velocities within the two distance scales $r_{xy}$ and $r_z$ that characterize the extension of the dark matter profile within the galactic planes and away from them. Prolate halos were actually observed in simulations of collisionless CDM~\cite{Dubinski:1991bm}, and were conjectured to result from either halo merging~\cite{Allgood:2005eu} or hierarchical structure formation, resulting in the collapse of matter along filaments rather than sheets~\cite{Bett:2006zy}. However, DM self-interactions tend to favor rounder halo shapes~\cite{Yoshida:2000uw}, and the MOND framework can only mimic spherical or slightly oblate DM distributions~\cite{Read:2005if}. As a result, ascertaining the actual presence of prolate halos would also provide additional evidence to discriminate among different scenarios, some of which also entail problematic features~\cite{Boran:2017rdn,kirill,Pardo:2020epc}.

In Section~\ref{sec:Qualitative_model} we elaborate on analytical and numerical computations of gravitational potentials created by a thin wire at the center of the galaxy, and also on those produced by dark--matter distributions elongated away from galactic planes with deformed Burkert~\cite{Burkert:1995yz} and NFW profiles~\cite{Navarro:1996gj}. In Section~\ref{sec:Data_fits}, we make use of the thin--wire potentials to fit a fair number of rotation curves from the SPARC database~\cite{Lelli:2016zqa}~\footnote{A number of Groningen Ph.D. Theses were instrumental to build the catalogue. More details can be found in~\cite{Lelli:2016zqa}.}; we then select two galaxies that demonstrate the most sizable improvement upon the addition of the wire, and test them with an alternative model involving deformed dark halos. We conclude in Section~\ref{sec:conclusions} with a summary of this work, some comments on its potential implications and a discussion of possible future lines of development.

\section{Deformed Dark--Matter Distributions}\label{sec:Qualitative_model}

Let us begin to explore the behavior of elongated matter distributions in the simplest possible setting, the admittedly academic problem of a massive wire of finite length $2 \ell_0$ through the center of a galaxy and orthogonal to its plane. Outside a spherical mass distribution, as is well known, the potential would have the standard $\frac{1}{r}$ behavior, while in this case
\begin{equation}
V\left(r\right) \ = \ G\,\mu \int_{-\,\ell_0}^{\ell_0} \frac{dz}{\sqrt{r^2+z^2}} \ = \ G\,\mu\,\ln\left(\frac{\sqrt{\ell_0^2+r^2}+\ell_0}{\sqrt{\ell_0^2+r^2}-\ell_0}\right) \ , \label{line_potential}
\end{equation}
where $\mu$ is a mass per unit length and $\ell_0$ is the length of the massive wire,
which exhibits an interesting transition between a $\log$--like behavior for $r\ll \ell_0$ and the standard $\frac{1}{r}$ behavior for $r\gg \ell_0$. In the former region the elongated mass distribution flattens the field lines toward the galactic plane, mimicking a two--dimensional log--like behavior, before the standard monopole term eventually dominates at large distances. As a result, the corresponding velocity distribution,
\begin{equation}\label{Wire_vel}
v_{DM}^2(r) \ = \ -\,r\,\frac{\partial V\left(r\right)}{\partial r} \ = \ \frac{2\,G\,\mu\, \ell_0}{\sqrt{\ell_0^2+r^2}} \ ,
\end{equation}
encodes a smooth transition between flat rotation curves and standard decaying profiles for $r \gg \ell_0$.
Our aim in the following is to ascertain whether, and to which extent, matter distributions of this type can account for flattened rotation curves. The potential interest in these considerations lies in the fact that they rest solely on the standard laws of gravity, and in particular on its Newtonian limit, without the need for any infrared modifications.

Let us turn to illustrate the other option that we are after, namely that the dark halo itself is not spherical, but is deformed in the direction orthogonal to the galactic plane. The Navarro-White-Frenk (NFW)~\cite{Navarro:1996gj} and Burkert~\cite{Burkert:1995yz} profiles,
\begin{equation}\label{Profiles}
\rho_{NFW}(r) \ = \ \frac{\rho_0r_0^3}{r\left(r+r_0\right)^2} \ ,
\qquad 
\rho_{B}(r) \ = \ \frac{\rho_0r_0^3}{\left(r^2+r_0^2\right)\left(r+r_0\right)} \ ,
\end{equation}
are among the most popular spherically symmetric smooth distributions used for dark matter in galaxies. Making the simple replacement
\begin{equation}
r \ = \ \sqrt{x^2+y^2+z^2} \ \to \ \sqrt{x^2+y^2+q\,z^2} \ , \label{deformed_r}
\end{equation}
one can deform these spherical distributions into prolate ones for $q<1$, or oblate ones for $q>1$. 
The resulting contributions to the rotation velocity are determined by
\begin{equation}
-\frac{\partial V}{\partial r} \ = \ 2G\int_0^\infty \ dz \ \int^\infty_0 \ dr'r'\rho \left(\sqrt{r'^2+q^2z^2}\right) \ \int_0^{2\pi} \  \frac{d\phi \left(r-r'\cos\phi\right)}{(r'^2+r^2+z^2-2rr'\cos\phi)^{3/2}} \ ,
\end{equation}
and performing the angular integral one can cast them in the form
\begin{equation}
\begin{gathered}\label{VDM}
v^2_{DM}(r) \ = \ -r\,\frac{\partial V}{\partial r} \ = \ 2G\int_0^\infty dz \ \int^\infty_0 dr'\,\frac{r'\rho \left(\sqrt{r'^2+q^2z^2}\right)}{\sqrt{(r-r')^2+z^2}}\left[F\left(\pi \ \Big| \ -\frac{4rr'}{(r-r')^2+z^2}\right)\right.\\
\left.\ - \ \frac{r'^2+z^2-r^2}{(r+r')^2+z^2}\,E\left(\pi\Big|-\frac{4rr'}{(r-r')^2+z^2}\right)\right] \ ,
\end{gathered}
\end{equation}
where 
\begin{equation}
F\left(\phi \ | \ k^2\right) \ = \ \int_0^\phi \ \frac{d\theta}{\sqrt{1-k^2\sin^2\theta}} \ ,\qquad 
E\left(\phi \ | \ k^2 \right) \ = \ \int_0^\phi \ d\theta \ \sqrt{1-k^2\sin^2\theta} 
\end{equation}
are incomplete elliptic integrals of the first and second kind.
\begin{figure}[ht]
\centering
	\includegraphics[width=90mm]{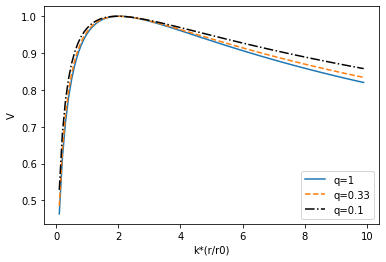}
	\caption{Dark--matter contributions to the rotation velocity for the NFW profile produced by a spherical halo (blue solid line), a prolate halo with a major-to-minor axis ratio 3 (orange dashed line), and a prolate halo with a major-to-minor axis ratio 10 (black dash-dotted line). The velocity is always normalized to unity at the peak, and the radius is normalized to $\frac{r_0}{k}$, with the coefficient $k$ always chosen so that the peak lies at $k\big(\frac{r}{r_0}\big)=2$. The actual value of $r_0$ depends on the galaxy.}
	\label{fig:NFW}
\end{figure}
Eq.~\eqref{VDM} is a complicated expression, which becomes however far simpler in the standard spherical limit ($q\rightarrow1$), and also in the limit of infinite elongation away from the galactic plane ($q\rightarrow0$). 
In the former case, the DM rotation velocity is determined by
\begin{equation}
v^2_{DM}(r) \ = \ \frac{4\pi G}{r}\int_0^r \ dr' \ r'^2 \rho(r') \ ,
\end{equation}
which results in
\begin{equation}\label{Vsph}
\begin{gathered}
v^2_{NFW}(r) \ = \ \frac{4\pi G\rho_0r_0^3}{r}\left[\ln(1+\frac{r}{r_0})-\frac{r}{r+r_0}\right] \ ,\\ 
v^2_B(r) \ = \ \frac{\pi G\rho_0r_0^3}{r}\left[\ln\left(1+\frac{r^2}{r_0^2}\right)+2\log\left(1+\frac{r}{r_0}\right)-2\arctan\left(\frac{r}{r_0}\right)\right]
\end{gathered}
\end{equation}
for the NFW and Burkert profiles.
\begin{figure}[ht]
\centering
	\includegraphics[width=90mm]{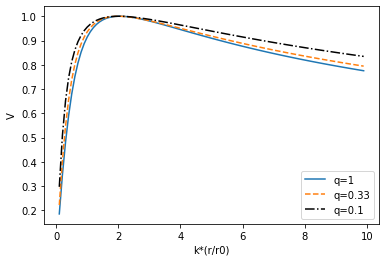}
	\caption{Dark--matter contributions to the rotation velocity for the Burkert profile produced by a spherical halo (blue solid line), a prolate halo with a major-to-minor axis ratio 3 (orange dashed line), and a prolate halo with a major-to-minor axis ratio 10 (black dash-dotted line). The velocity is always normalized to unity at the peak, and the radius is normalized to $\frac{r_0}{k}$, with the coefficient $k$ always chosen so that the peak lies at $k\big(\frac{r}{r_0}\big)=2$. The actual value of $r_0$ depends on the galaxy.}
	\label{fig:Burkert}
\end{figure}
On the other hand, in the limit of infinite elongation the rotation velocity reflects the two--dimensional version of Gauss's theorem, so that
\begin{equation}
v^2_{DM}(r) \ = \ 4\pi G \int_0^r \ dr' \ r'\rho(r') \ ,
\end{equation}
where $r=\sqrt{x^2+y^2}$. In detail, for the deformed NFW and Burkert profiles
\begin{equation}
\begin{gathered}
v^2_{NFW}(r) \ = \ 4\pi G\rho_0r^2_0\, \frac{r}{r+r_0} \ ,\\
v^2_B(r) \ = \ \pi G\rho_0r^2_0\left[\ln\left(1+\frac{r^2}{r_0^2}\right)-2\ln\left(1+\frac{r}{r_0}\right)+2\arctan\left(\frac{r}{r_0}\right)\right] \ .
\end{gathered}
\end{equation}

From these expressions one can see that, if the density profile $\rho$ has a characteristic radial scale $r_0$ beyond which it tends smoothly to zero, eventually $v^2_{NFW, B}\propto r^{-1}\ln{r}$ in the spherical limit but $v^2_{NFW,B}\propto {\mathrm{const}}$ in the limit of infinite elongation. Even in the presence of smooth mass distributions, the effect we are after can therefore grant the emergence of flat rotation curves for distances scales $r_0 < r < q^{-1}r_0$, if $q<1$.
On the other hand, inside dark matter distributions one would observe growing rotation curves in both limits, with $v_{NFW}^2\propto r$ and $v_B^2\propto r^2$ for NFW and Burkert profiles. For generic bulged profiles the integrals must be computed numerically, and figs.~\ref{fig:NFW} and \ref{fig:Burkert} compare the results thus obtained with the spherical limits for the two cases of NFW and Burkert profiles.
Our conclusion if therefore that, \emph{for both NFW and Burkert profiles the dark--matter contributions to rotation velocities exhibit steep rises followed by shallow declines, and prolate halos yield steeper rises and shallower declines than the standard spherical ones}. This behavior in the outer region complies to the picture that emerged from the case of a thin wire. However, we expect the effect to be considerably less significant in this case, since physically realistic values of $q$ are not expected to be smaller than $0.33-0.5$.

Before starting the quantitative analysis, it is useful to introduce a different parametrization of velocities and density profiles, which rests on the definition of the virial mass of the halo:
\begin{equation}
M \ = \ \int d^3r \ \rho(\vec{r}) \ = \ \frac{4\pi}{q}\int_0^{r_c} dr \ r^2\rho(r) \ .
\end{equation}
Here we have defined $r_c$, in an admittedly arbitrary but reasonable manner, demanding that the average DM density within the ellipsoid be 200 times larger than the critical density of the Universe~\cite{White:2000jv}:
\begin{equation}\label{Mcr}
M \ = \ 200 \ * \ \frac{4}{3}\pi\frac{\rho_cr_c^3}{q} \ = \ 100\frac{H^2}{qG}r_c^3 \ ,
\end{equation}
where we have taken the Hubble parameter $H$ to be $0.072$ km/s/kpc. For the NFW and Burkert profiles, the halo masses are
\begin{equation}
\begin{gathered}
M_{NFW} \ = \ \frac{4\pi\rho_0r_0^3}{q} \left[\ln(1+\frac{r_c}{r_0})-\frac{r_c}{r_c+r_0}\right] \ , \\
M_B \ = \ \frac{\pi\rho_0r_0^3}{q}\left[\ln(1+\frac{r_c^2}{r_0^2})+2\ln(1+\frac{r_c}{r_0})-2\arctan(\frac{r_c}{r_0})\right] \ .
\end{gathered}
\end{equation}
Equating these expressions to~\eqref{Mcr} gives
\begin{equation}\label{NFW}
\rho_0 \ = \ \frac{25H^2}{\pi G}\frac{C^3}{\ln(1+C)-\frac{C}{1+C}}
\end{equation}
for the NFW case, and
\begin{equation}\label{B}
\rho_0 \ = \ \frac{100H^2}{\pi G}\frac{C^3}{\ln(1+C^2)+2\ln(1+C)-2\arctan(C)}
\end{equation}
for the Burkert case, where the parameter
\begin{equation}
C \ = \ \frac{r_c}{r_0}
\end{equation}
is known as concentration. For $q=1$, one can easily plug this expression into~\eqref{Vsph} to obtain that the rotation velocity at $r_c$ is given by
\begin{equation}
v_c \ = \ 10Hr_0C \ 
\end{equation}
for \emph{any} spherical DM profile. We can now rewrite~\eqref{Vsph} in terms of the new parameters, $C$ and $v_c$:
\begin{equation}
\label{NFWB}
\begin{gathered}
v^2_{NFW}(r) \ = \ v_c^2 \frac{v_c}{10Hr}\frac{\ln(1+10HCv^{-1}_cr)-10HCv^{-1}_cr(1+10HCv^{-1}_cr)^{-1}}{\ln(1+C)-C(1+C)^{-1}} \ , \\
\\
v^2_B(r) \ = \ v_c^2 \frac{v_c}{10Hr}\frac{\ln(1+(10HCv^{-1}_cr)^2)+2\ln(1+10HCv^{-1}_cr)-2\arctan(10HCv^{-1}_cr)}{\ln(1+C^2)+2\ln(1+C)-2\arctan(C)} \ .
\end{gathered}
\end{equation}
Likewise, for very small values of $q$ close to zero
\begin{equation}
v^2_{NFW}(r) \ \approx \ v_c^2\left(\frac{1+C}{C}\right)\left(\frac{r}{r+r_0}\right) \ ,
\end{equation}
with
\begin{equation}
r_0 \ = \ \frac{v_c}{10HC^2}\sqrt{(1+C)\ln(1+C)-C} \ ,
\end{equation}
and
\begin{equation}
v_B^2(r) \ \approx \ v_c^2\left(\frac{\ln\left(1+\frac{r^2}{r_0^2}\right)-2\ln\left(1+\frac{r}{r_0}\right)+2\arctan\left(\frac{r}{r_0}\right)}{\ln\left(1+C^2\right)-2\ln\left(1+C\right)+2\arctan\left(C\right)}\right) \ ,
\end{equation}
with
\begin{equation}
r_0 \ = \ \frac{v_c}{10H}\left(\frac{\ln(1+C^2)+2\ln(1+C)-2\arctan(C)}{\ln(1+C^2)-2\ln(1+C)+2\arctan(C)}\right)^{1/2} \ .
\end{equation}
For generic values of $q$, the expression for rotation velocity would be more complicated,
\begin{equation}
\label{vdef}
v^2(r,v_c,C,q) \ = \ v_c^2\frac{F\left(\sqrt{2G\rho_0F(C,q)}v_c^{-1}r,q\right)}{F(C,q)} \ ,
\end{equation}
where $F$ is the function
\begin{equation}
\begin{gathered}
F(x,q) \ = \ q^{-1} \int_0^\infty d\xi \ \xi^2\bar{\rho} (\xi) \ \int^{\pi/2}_0 \frac{d\theta\, \cos{\theta}}{\sqrt{(x-\xi\cos\theta)^2+(\xi/q)^2\sin^2\theta}}\left[F\left(\pi \ \Big| u\right)\right.\\
 \\
\left.\ - \ \frac{\xi^2\cos^2\theta+(\xi/q)^2\sin^2\theta-x^2}{(x+\xi\cos\theta)^2+(\xi/q)^2\sin^2\theta}\,E\left(\pi\Big|u\right)\right]    \ ,
\end{gathered}
\end{equation}
with
\begin{equation}
u \ = \ -\frac{4x\xi\cos\theta}{(x-\xi\cos\theta)^2+(\xi/q)^2\sin^2\theta} \ , \qquad \bar{\rho}(x) \ = \ \frac{\rho(xr_0)}{\rho_0} \ ,
\end{equation}
and $\rho_0$ is either~\eqref{NFW} or~\eqref{B}, depending on the choice of profile.
\section{Fits of Observational Data}\label{sec:Data_fits}
In order to select a suitable sample of galaxies from the SPARC database, we have resorted to the following criteria:
\begin{itemize}
  \item The galaxy should have the quality flag $Q=1${, which means high quality HI (the 21-cm line from neutral atomic hydrogen) or hybrid HI/H$\alpha$ (the strongest emission line of ionized hydrogen) data}~\cite{Lelli:2016zqa};
  \item The galaxy's inclination should be equal to or larger than 30${}^o$~\cite{Lelli:2016zqa};
  \item The number of data points should be equal to or larger than 10. This requirement is admittedly arbitrary, but if the length of the curve is too small, identifying the ``flattening`` effect we are after and discriminating among competing models becomes very difficult. A similar choice was made in an earlier work~\cite{Bondarenko:2020mpf};
\end{itemize}
These criteria leave 84 galaxies out of 175, on which the deformations we intend to explore are expected to lead to more sizable effects, if they are relevant at all to them.

We then performed a five--parameter fit with the function
\begin{equation}
\small
\label{Vstring}
v(r; Y_D,Y_B,v_c,C,\mu)=\sqrt{Y_Dv^2_D(r)+Y_Bv_B^2(r)+v_G(r)|v_G(r)|+v^2_{DM}(r,v_c,C)+2G\mu} \ ,
\end{equation} 
using the Python \textit{SciPy.Optimize.Curve\textunderscore Fit} package. The five parameters are the mass-to-light ratios $Y_D$ and $Y_B$, bounded from below at 0.1, the two parameters $v_c$ and $C$ of the NFW and Burkert profiles of eq.~\eqref{NFWB}, and finally the string tension parameter $\mu$. The disk, bulge, and gas contributions $v_{D,B,G}$ are taken from the SPARC database; notice that the absolute value is needed for the gas contribution, because it can become negative at small radii if the gas distribution is significantly depressed in the innermost regions, so that the gravitational pull from outwards is stronger than from inwards~\cite{Lelli:2016zqa}. The galaxy UGC 01281 has particularly large negative values of $v_G$ at small radii, rendering the fit unstable; for this reason, we excluded it from the sample, confining our attention to a total of 83 objects. Given that each object was fitted by two profiles (NFW and Burkert), this means 166 models in total, which is indeed a large number.

Then, taking the fit results as a starting point, we performed an MCMC analysis on them resorting to the Python package \textit{emcee}~\cite{ForemanMackey:2012ig}. As in~\cite{Li:2020iib}, we imposed flat priors on $V_c$ and $C$, with $10<V_c<500$ and $0<C<1000$. We also included two additional parameters, namely the inclination of the galaxy $i$ and the distance to the galaxy $D$, and imposed Gaussian priors on them, with their mean values and standard deviations taken from the SPARC database. The former affects the observed velocities and their errors as
\begin{equation}\label{Inc}
v_{obs}' \ = \ \frac{sin(i)}{sin(i')}\,v_{obs} \ , \ \delta v_{obs}' \ = \ \frac{sin(i)}{sin(i')}\,\delta v_{obs} \ ,   
\end{equation}
while the latter impacts the disk, gas, and bulge components:
\begin{equation}
v'_{d,b,g} \ = \ \sqrt{\frac{D'}{D}}\, v_{d,b,g} \ .   
\end{equation}
We chose our posterior probability to be proportional to $e^{-L}$, with
\begin{equation}
\begin{gathered}
L \ = \ \frac{1}{2}\chi^2 \ + \ \frac{1}{2}\left(\frac{\log_{10}(Y_D)-log_{10}(0.5)}{0.1}\right)^2 \ + \ \frac{1}{2}\left(\frac{\log_{10}(Y_B)-log_{10}(0.7)}{0.1}\right)^2 \\
\ + \ \frac{1}{2}\left(\frac{D-D_0}{\delta D}\right)^2 \ + \ \frac{1}{2}\left(\frac{i-i_0}{\delta i}\right)^2 \ ,
\end{gathered}
\end{equation}
with $\chi$-square given by
\begin{equation}
\chi^2 \ = \ \sum_i\left(\frac{v(x_i)-v_{obs}(x_i)}{\delta v_{obs}(x_i)}\right)^2 \ ,\\
\end{equation}
and
\begin{equation}
v(r; Y_D,Y_B,v_c,C,i,D)=\frac{sin(i)}{sin(i_0)}\sqrt{(D/D_0)\left(Y_Dv^2_D(r)+Y_Bv_B^2(r)+v_G(r)|v_G(r)|\right)+v^2_{DM}(r,v_c,C)} \ 
\end{equation}
(the prefactor $\frac{sin(i)}{sin(i_0)}$ is due to~\eqref{Inc}; we multiplied both the numerator and the denominator in $\chi^2$ by $\frac{sin(i_0)}{sin(i)}$).

The first term in $L$ is the likelihood function, and all the rest are priors (we omit the normalization factor, since it is irrelevant for our purpose). Namely, as in~\cite{Li:2020iib}, we imposed Gaussian priors on $D$ and $i$, with the mean values and errors given by the SPARC database, and lognormal priors on $Y_D$ and $Y_B$, with mean values of $0.5$ and $0.7$ and error values of $0.1$. The third term, which has to do with bulge, is omitted for bulgeless galaxies.

Following~\cite{Li:2020iib}, we initialized the MCMC chains with 200 random walkers and ran 500 burn-in iterations, before resetting the sampler and running another 2000 iterations. For fits that have a minimum at $G\mu=0$, where $G \mu$ is the tension term in eq.~\eqref{Vstring}, we ran only a 6-parameter MCMC ($v_c, C, Y_D, Y_B, D$, and $i$), while for those that have a minimum at $G\mu>0$ (69 in total), we also performed a 7-parameter MCMC with $G\mu$ as the seventh parameter. Then, using the obtained values of $L$, we computed the Bayesian and Akaike information criteria (BIC and AIC) to compare the evidence for competing models. BIC is given by
\begin{equation}
BIC \ = \ k\ln(n)+2L \ , 
\end{equation}
where $k$ is the number of parameters (6 and 7 for bulged and bulgeless galaxies, respectively, when the filament is present, and 5 and 6, respectively, when the filament is absent), and $n$ is the number of data points. Likewise, AIC is defined as 
\begin{equation}
AIC \ = \ 2k+2L \ .
\end{equation}
For cases when non--zero values of $G\mu$ yielded improvement of BIC and/or AIC, we also considered a more physically realistic situation when the filament at the center of the galaxy has finite length of 200 kpc, which means that the factor of $2G\mu$ in~\eqref{Vstring} was replaced with $2G\mu\left(1+(r/100\right)^2)^{-1/2}$, as per~\ref{Wire_vel}. Based on the result, we selected 23-25 galaxies that have \textit{some} evidence for a ``wire`` at the center (values of fit parameters are given in Tables \ref{tab:table1},~\ref{tab:table2},~\ref{tab:table3}), and split them into three groups: those with strong evidence, when considerable fit improvement is shown for both NFW and Burkert profiles, implying a model--independent feature (9 galaxies for BIC, 11 for AIC; Tables ~\ref{tab:table4} and~\ref{tab:table8}); those with moderate evidence, where the addition of the filament improves fit quality for the better--fitting profile (6 galaxies for BIC, 7 for AIC; Tables~\ref{tab:table5} and~\ref{tab:table9}); those with scant evidence, where the filament improves the fitting quality for the worse--fitting profile (8 galaxies for BIC, 7 for AIC; Tables~\ref{tab:table6} and~\ref{tab:table10}). The values of BIC and AIC for the galaxies that show~\textit{no} improvement upon the addition of the filament are given in Tables~\ref{tab:table9} and~\ref{tab:table11}.

\begin{figure}[ht]
\centering
\begin{adjustbox}{width={\textwidth},totalheight={\textheight},keepaspectratio}
\begin{tabular}{cc}
		\includegraphics[width=75mm]{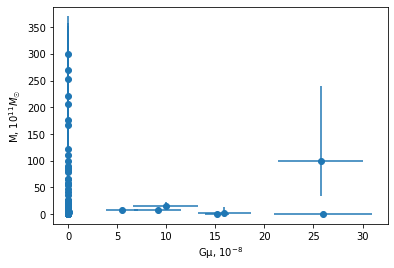} &
		\includegraphics[width=75mm]{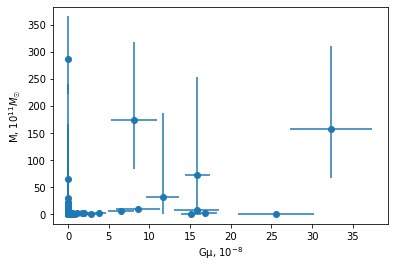} \\
	\end{tabular}
\end{adjustbox}
	\caption{Left panel: string tension versus dark halo mass for the NFW profiles; right panel: string tension versus dark halo mass for the Burkert profiles. In the NFW model, there is no obvious correlation between the presence of a string and larger halo masses, while in the Burkert model, one can observe some degree of correlation. The asymmetric error bars on the masses are due to the fact that the masses are proportional to $v_c^3$, and $(v_c-\delta v_c)^3$ and $(v_c+\delta v_c)^3$ are not equidistant from $v_c^3$.}
		\label{MassString}
\end{figure}

The nature of these string--like objects remains unclear however, and one could conjecture that they originate from intergalactic filaments of the large--scale structure. In this case, there would be a correlation between larger masses of the dark halos and the presence of a ``string`` at the center, with the largest galaxies located at the nodes of the filaments. However, based on our data sample, we cannot reach a clear conclusion on whether or not such correlation exists: there appear to be some indications to this effect in the Burkert model, while they do not emerge in the NFW model (Fig.~\ref{MassString}).

\begin{figure}[ht]
\centering
\begin{adjustbox}{width={\textwidth},totalheight={\textheight},keepaspectratio}
\begin{tabular}{cc}
		\includegraphics[width=75mm]{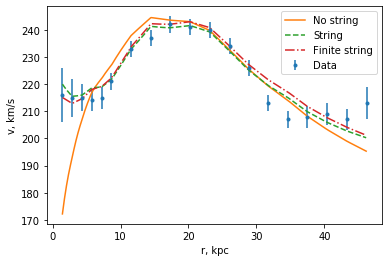} &
		\includegraphics[width=75mm]{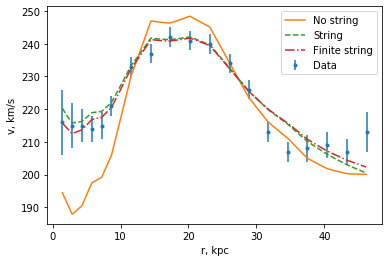} \\
	\end{tabular}
\end{adjustbox}
	\caption{Rotation curves for the galaxy NGC 5371: data points (blue dots with error bars), fit with a DM halo (solid orange line), a DM halo plus an infinite string-like object at the origin (dashed green line), and a DM halo plus a finite string-like object of length 200 kpc (dash-dotted red line). Left panel: NFW profile; right panel: Burkert profile.}
		\label{NGC5371}
\end{figure}

\begin{figure}[ht]
\centering
\begin{adjustbox}{width={\textwidth},totalheight={\textheight},keepaspectratio}
\begin{tabular}{cc}
		\includegraphics[width=75mm]{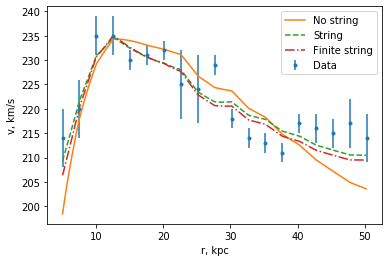} &
		\includegraphics[width=75mm]{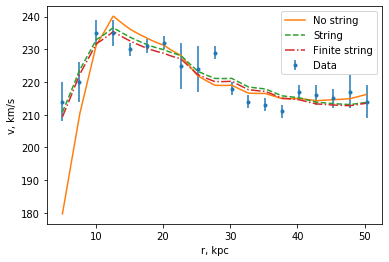} \\
	\end{tabular}
\end{adjustbox}
	\caption{Rotation curves for the galaxy NGC 5907: data points (blue dots with error bars), fit with a DM halo (solid orange line), a DM halo plus an infinite string-like object at the origin (dashed green line), and a DM halo plus a finite string-like object of length 200 kpc (dash-dotted red line). Left panel: NFW profile; right panel: Burkert profile.}
		\label{NGC5907}
\end{figure}

\begin{figure}[ht]
\centering
	\includegraphics[width=120mm]{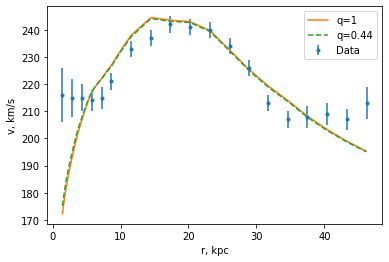}
	\caption{The rotation curve of NGC 5371, fitted by a spherical dark halo (orange solid curve) and by a deformed halo (green dashed curve). The result is given for the NFW profile.}
	\label{NGC5371_q_NFW}
\end{figure}

\begin{figure}[ht]
\centering
	\includegraphics[width=120mm]{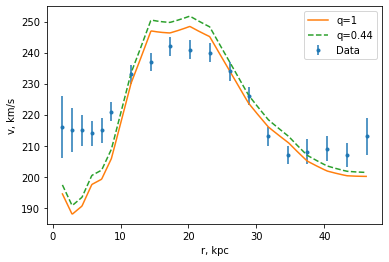}
	\caption{The rotation curve of NGC 5371, fitted by a spherical dark halo (orange solid curve) and by a deformed halo (green dashed curve). The result is given for the Burkert profile.}
	\label{NGC5371_q_B}
\end{figure}

\begin{figure}[ht]
\centering
	\includegraphics[width=120mm]{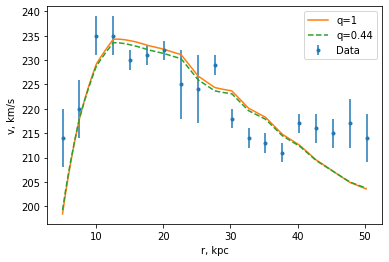}
	\caption{The rotation curve of NGC 5907, fitted by a spherical dark halo (orange solid curve) and by a deformed halo (green dashed curve). The result is given for the NFW profile (for the Burkert profile, a deformed halo is disfavored in comparison to the spherical one).}
	\label{NGC5907_q}
\end{figure}

Two most notable examples of galaxies belonging to the first category are NGC 5371 and NGC 5907, for both of which the fit improves considerably due to the presence of the filament (Figs.~\ref{NGC5371} and~\ref{NGC5907}). Though BIC is more stringent than AIC, the conclusions for both criteria are similar in most cases (the few counterexamples, such as NGC 3521 and UGC 03205, yield only marginal improvement in AIC that could be attributed to statistical uncertainty). Likewise, the addition of a cutoff does not have any significant impact on either BIC or AIC for most galaxies.
Taking the aforementioned galaxies NGC 5371 and NGC 5907 as benchmarks, we also fitted them with the alternative model that involves prolate dark matter halos. In this case, the rotation velocity would be
\begin{equation}
\small
\label{Vhalo}
v(r; Y_D,Y_B,v_c,C,\mu)=\sqrt{Y_Dv^2_D(r)+Y_Bv_B^2(r)+v_G(r)|v_G(r)|+v^2_{DM}(r,v_c,C,q)} \ ,
\end{equation}
with $v^2_{DM}$ given by~\eqref{vdef}. We opted for a flat prior on $q$ in the range from 1/3 to 1, i. e. between the limiting cases of a spherical halo and halo with a major-to-minor axis ratio of 3. The MCMC analysis results, along with the values of BIC and AIC, are given in Table~\ref{tab:table12}. As can be seen from these values and from the examples of the same two galaxies NGC 5371 and NGC 5907 (figs.~\ref{NGC5371_q_NFW},~\ref{NGC5371_q_B}, and~\ref{NGC5907_q}), the deformation of the halo gives only a small correction to the fit, unlike the filament at the center.

Finally, for the same two galaxies, we have explored the possibility that the halo deformation and the filament be simultaneously present. However, this model is not supported by the data analysis: for the NGC 5371 (both NFW and Burkert) and NGC 5907 NFW, the best fit obtains for $q=1$ once one incorporates both parameters $q$ and $G\mu$. For NGC 5907 with Burkert profile, there is a minimum at $q<1$, but the BIC and AIC results in this case are worse than for a filament without the deformation (the values of the fitting parameters, BIC and AIC for this case are given in Table~\ref{tab:table13}). We hope to return this this important issue with a more detailed analysis, which would require computer facilities beyond those presently at our disposal.
\section{Conclusions}\label{sec:conclusions}

In this paper we have elaborated on the idea that flattened rotation curves could reflect the presence of a long thin filament at the center of the galaxy or, alternatively, the elongation of dark matter halos away from the galactic plane. As we have seen in Section~\ref{sec:Qualitative_model}, \textit{a string--like object at the origin can directly produce a logarithmic term in the gravitational potential, while bulged dark--mater distributions yield quasi-logarithmic potentials within a certain range of distances by squeezing Newtonian field lines around galactic planes}. This setting would be at variance with the standard picture, which ascribes flattened rotation curves to the dominant presence of dark matter in the outer regions of galactic planes.

In the first part of this work, after highlighting the phenomenon in its simplest instance, when the logarithmic potential is produced by a thin wire, we have turned to the two popular NFW and Burkert density profiles, whose spherical versions are often used for fits of galaxy rotation curves. However, we have deformed these spherical profiles performing in their arguments the simple substitution in eq.~\eqref{deformed_r},
so that $q<1$ corresponds to bulged dark--matter profiles, while $q>1$ corresponds to dark--matter profiles that are squeezed around galactic planes. While admittedly arbitrary, this simple deformation is arguably a suggestive departure from spherical profiles that grants a first comparison of our picture with data at the price of a single additional parameter. Armed with these ingredients, we have performed fits of 83 galaxy rotation curves from the SPARC sample, selected on the basis of the quality and quantity of the available data on them, with spherical NFW and Burkert profiles profiles with a string--like filament at the origin, and discovered that the addition of the filament does improve the fit quality to varying degrees for about 28\% of the examined galaxies (23-25 galaxies out of the 83 that we had originally selected). We then chose two of them that demonstrated particular improvement in fit quality, and fitted them with alternative models of elongated mass distributions outside of the galactic plane, namely prolate dark halos. However, in this case the improvement within a physically plausible range of major-to-minor axis ratios (i.e. not larger than 3) was very modest at best, and in one case, the extra deformation parameter was disfavored by both Bayesian and Akaike criteria. We have also attempted to fit both galaxies with an extended model that incorporates both a deformed halo and a filament at the center, but the present indication is that the halo deformation appears disfavored for both galaxies and both profiles.

There are numerous physical motivations to assume the existence of elongated mass distributions at the centers of the galaxies. First and foremost, black holes can produce relativistic jets comprised of gas~\cite{elongated}, and given the gravitational nature of the effect, dark--matter jets can in principle exist as well. One of our primary examples, the galaxy NGC 5371, is classified as a LINER, which may imply the presence of an active galactic nucleus~\cite{Rush:1993qz}. Alternatively, such an object could be a tidal stream: for instance, another key example NGC 5907 is known to host an extended stellar tidal stream structure~\cite{MartinezDelgado:2008cx,Dokkum}. One more prominent example, NGC 2841, has a polar ring orthogonal to the galaxy plane, which is probably a result of interaction with another galaxy~\cite{Kaneda:2007uw,Afanasiev:1998xi}. In addition, one could conjecture that the ``strings`` at the centers of galaxies are connected to the large--scale intergalactic filaments. In this case, one ought to observe these objects more clearly in more massive galaxies. However, due to the limitations on the sample of examples at our disposal and, above all, to the large fitting errors on halo masses, our analysis is unable, at present, to come to a definite conclusion on a possible correlation between larger halo masses and elongated objects at the center. Finally, there are exotic candidates like cosmic strings: a mechanism for the migration of cosmic strings to the center of galaxies was proposed in~\cite{Vilenkin:2018zol}, and interestingly, the values of $G\mu$ obtained in our fits are below the upper constraint on cosmic string tension obtained from Planck observations, which is around $7.8*10^{-7}$~\cite{Ade:2013xla}. A filament of this type, observed at the center of Milky Way, has also been conjectured to be either a black hole jet or a cosmic string~\cite{morris}. The Milky Way hosts several more such objects, which have been interpreted as jets produced by shock waves within the regions of intensive star formation~\cite{YusefZadeh:2003qx}.

Likewise, if prolate halos were to prove a generally dominant feature, this would have some bearing on the controversy between the cold dark matter (CDM) paradigm, self-interacting dark matter (SiDM), and models of modified gravity or modified Newtonian dynamics (MOND). To wit, CDM simulations often produce elongated halos~\cite{Dubinski:1991bm}, while SiDM favors rounded halos~\cite{Yoshida:2000uw}, and MOND is only expected to imitate spherical or slightly oblate dark density distributions~\cite{Read:2005if}. It is not clear to us at present whether a sharper evidence for prolate halos could also impinge on hybrid models, such as superfluid dark matter, which at finite temperatures is expected to yield both a MOND-like modification of gravity and a CDM-like matter component~\cite{Berezhiani:2015bqa}.

This work has clearly some limitations, which make its conclusions at best preliminary. Most notably, we adopted the simplest possible model of the elongated mass distribution at the center, that of an infinitely thin wire. We can only justify it on grounds of economy of parameters and variety of the scenarios that it can approximate. Nonetheless, we found that this model significantly improves the quality of the fits for about 28\% of the 83 galaxies that we examined in detail. This result lends, in our opinion, some credence to the dynamical effect we were after, which would also forego the need for infrared modifications of gravity in this context.

Likewise, when analyzing prolate halos, we have worked with simple deformations of two specific, if very popular, density profiles, and confined our analysis to only two galaxies. The prefatory conclusion, based on both theoretical considerations and fit results for these two examples, is that the improvement in the fit quality due to halo deformation is marginal at best. This implies that the rotation curve analysis is not conclusive, and may probably be more useful as an \textit{exclusionary} method to filter out the galaxies that are unlikely to host non--spherical dark halos. Observations of gravitational lensing indicate that some galaxies and galaxy clusters have prolate dark matter distributions~\cite{Hoekstra:2003pn,Oguri:2004in}, and a full--fledged analysis of all 83 galaxies, which will be performed in a future work, will be instrumental to reinforce or disprove the present findings. 

In principle, the galaxies that exhibit~\textit{some} preference for prolate shapes, especially when the indication obtains for both NFW and Burkert profiles, can be tested independently by other observations, and in particular by the detection of kinematic stellar groups~\cite{RojasNino:2011xh,Rojas-Nino:2015qna}. The more refined galaxy surveys expected from the Euclid mission are likely to shed more light on these important issues~\cite{Laureijs:2011gra}.


\acknowledgments %
I am grateful to S.~McGaugh, F.~Lelli, and P.~Li for clarifications on the SPARC database, to Ivano Basile for discussions and help with Cython codes, and to Andrea Pallottini for technical assistance. I am very grateful to A.~Sagnotti for suggesting to look into halo bulges away from galactic planes and to A.~Ferrara for his detailed comments and suggestions on the manuscript.
I am also grateful to DESY-Hamburg, where part of this work was done, for the kind hospitality extended to me. Finally, I would like to thank the referee Pier-Stefano Corasaniti for his constructive (and instructive) criticism of an earlier version of this work. This reserach was supported in part by Scuola Normale Superiore, by INFN (IS CSN4-GSS-PI) and by the MIUR-PRIN contract 2017CC72MK\_003.


\bibliographystyle{JCAP}

\newpage
\footnotesize
\begin{longtable}{|| c || c | c | c | c | c | c |} 
 \centering
    Galaxy & $V_{200}$, km/s & $C_{200}$ & $Y_d$ & $Y_b$ & D, Mpc & i, $^o$ \\ \hline\hline
    D631-7, NFW & 372.9$\pm$88 & 0.6$\pm$0.45 & 0.33$\pm$0.07 & 0 & 7.62$\pm$0.18 & 53.65$\pm$3.04 \\ \hline
    D631-7, B & 65.04$\pm$8.74 & 12.86$\pm$0.87 & 0.42$\pm$0.09 & 0 & 7.68$\pm$0.18 & 56.74$\pm$2.96 \\ \hline
    DDO064, NFW & 224.07$\pm$176.15 & 1.69$\pm$2.22 & 0.53$\pm$0.14 & 0 & 8.02$\pm$2 & 60.69$\pm$4.94 \\ \hline
    DDO064, B & 39$\pm$41.51 & 22.43$\pm$3.9 & 0.55$\pm$0.15 & 0 & 7.34$\pm$2.09 & 60.33$\pm$5 \\ \hline
    DDO161, NFW & 99.65$\pm$21.74 & 2.23$\pm$0.82 & 0.36$\pm$0.08 & 0 & 7.94$\pm$1.5 & 69.91$\pm$8.96 \\ \hline
    DDO161, B & 55.19$\pm$3.34 & 10.26$\pm$1.11 & 0.52$\pm$0.12 & 0 & 7.78$\pm$1.58 & 69.67$\pm$9.5 \\ \hline
    ESO079-G014, NFW & 304.46$\pm$116.15 & 2.62$\pm$1.34 & 0.63$\pm$0.12 & 0 & 32.61$\pm$5.78 & 79.13$\pm$4.77 \\
\hline
    \end{longtable}
\newpage
\begin{longtable}{|| c || c | c | c | c | c | c | c |} 
 \centering
    Galaxy & $V_{200}$, km/s & $C_{200}$ & $Y_d$ & $Y_b$ & D, Mpc & i, $^o$ \\ \hline\hline
    NGC 2955, B & 159.43$\pm$13.18 & 19.86$\pm$2.59 & 0.34$\pm$0.07 & 0.98$\pm$0.16 & 94.23$\pm$9.23 & 50.74$\pm$5.47 \\ \hline
    NGC 2998, NFW & 136.78$\pm$6.7 & 14.67$\pm$3.86 & 0.42$\pm$0.12 & 0 & 65.11$\pm$10.3 & 57.81$\pm$2 \\ \hline
    NGC 2998, B & 141.72$\pm$9.22 & 9.1$\pm$1.24 & 0.73$\pm$0.1 & 0 & 78.03$\pm$8.64 & 58.6$\pm$1.94 \\ \hline
    NGC 3109, NFW & 428.44$\pm$50.86 & 0.48$\pm$0.23 & 0.43$\pm$0.1 & 0 & 1.38$\pm$0.07 & 74.03$\pm$4.45 \\ \hline
    NGC 3109, B & 56.49$\pm$3.18 & 16.65$\pm$0.75 & 0.53$\pm$0.14 & 0 & 1.33$\pm$0.07 & 69.67$\pm$4.97 \\ \hline
    NGC 3198, NFW & 110.52$\pm$2.44 & 10.46$\pm$0.8 & 0.46$\pm$0.06 & 0 & 13.58$\pm$1.32 & 72.72$\pm$3.09 \\ \hline
    NGC 3198, B & 99.94$\pm$2.03 & 15.04$\pm$1.12 & 0.66$\pm$0.07 & 0 & 14.61$\pm$1.28 & 73.56$\pm$2.89 \\
\hline
            \end{longtable}
\newpage
\renewcommand\thetable{1}
\begin{longtable}{|| c || c | c | c | c | c | c | c |} 
 \centering
    Galaxy & $V_{200}$, km/s & $C_{200}$ & $Y_d$ & $Y_b$ & D, Mpc & i, $^o$ \\ \hline\hline
    NGC 6195, B & 182.02$\pm$35.68 & 9.92$\pm$2.06 & 0.42$\pm$0.08 & 0.79$\pm$0.1 & 126.13$\pm$11.68 & 61.12$\pm$4.72 \\ \hline
    NGC 6503, NFW & 80.77$\pm$1.55 & 14.44$\pm$0.97 & 0.4$\pm$0.04 & 0 & 6.15$\pm$0.3 & 73.67$\pm$2.04 \\ \hline
    NGC 6503, B & 74.01$\pm$1.01 & 20.6$\pm$0.77 & 0.57$\pm$0.04 & 0 & 6.25$\pm$0.3 & 73.91$\pm$2.01 \\ \hline
    NGC 6674, NFW & 409.98$\pm$62.3 & 0.56$\pm$0.37 & 0.82$\pm$0.12 & 0.88$\pm$0.22 & 65.14$\pm$7.78 & 60.61$\pm$4.5 \\ \hline
    NGC 6674, B & 258.74$\pm$35.11 & 4.25$\pm$0.39 & 0.89$\pm$0.14 & 0.76$\pm$0.18 & 68.39$\pm$8.27 & 61.24$\pm$5.14 \\ \hline
    NGC 6946, NFW & 117.59$\pm$10.15 & 10.21$\pm$2.09 & 0.5$\pm$0.08 & 0.56$\pm$0.09 & 5.05$\pm$0.93 & 37.63$\pm$1.93 \\ \hline
    NGC 6946, B & 90.83$\pm$6.65 & 19.7$\pm$5.06 & 0.57$\pm$0.09 & 0.65$\pm$0.11 & 5.06$\pm$0.88 & 37.52$\pm$1.96 \\
\hline
    \end{longtable}
\newpage
\renewcommand\thetable{1}
\begin{longtable}{|| c || c | c | c | c | c | c | c |} 
 \centering
    Galaxy & $V_{200}$, km/s & $C_{200}$ & $Y_d$ & $Y_b$ & D, Mpc & i, $^o$ \\ \hline\hline
    UGC 06983, NFW & 79.09$\pm$6.44 & 12.7$\pm$2.06 & 0.52$\pm$0.13 & 0 & 17.76$\pm$2.47 & 48.96$\pm$1 \\ \hline
    UGC 06983, B & 63.51$\pm$2.6 & 27.05$\pm$2.45 & 0.54$\pm$0.14 & 0 & 18.33$\pm$2.41 & 49$\pm$1 \\ \hline 
    UGC 07125, NFW & 55.55$\pm$5.34 & 5.15$\pm$0.92 & 0.52$\pm$0.13 & 0 & 14.28$\pm$4.82 & 88.02$\pm$1.38 \\ \hline
    UGC 07125, B & 43.05$\pm$2.67 & 12.41$\pm$1.32 & 0.53$\pm$0.13 & 0 & 17.42$\pm$4.13 & 87.94$\pm$1.45 \\ \hline
    UGC 07151, NFW & 68.2$\pm$38.11 & 6.95$\pm$2.96 & 0.8$\pm$0.17 & 0 & 7$\pm$0.34 & 87.98$\pm$1.41 \\ \hline
    UGC 07151, B & 37.02$\pm$1.73 & 31.77$\pm$2.83 & 0.56$\pm$0.14 & 0 & 6.87$\pm$0.34 & 87.98$\pm$1.42 \\ \hline
    UGC 07323, NFW & 300.09$\pm$132.71 & 1.51$\pm$1.16 & 0.48$\pm$0.11 & 0 & 8.43$\pm$2.4 & 47.13$\pm$3 \\
\end{longtable}
\newpage
\begin{longtable}{|| c || c | c | c | c | c | c | c |} 
 \centering
    Galaxy & $V_{200}$, km/s & $C_{200}$ & $Y_d$ & $Y_b$ & D, Mpc & i, $^o$ & $G\mu, 10^{-8}$  \\ \hline\hline
   DDO064, B & 46.64$\pm$135.33 & 20.49$\pm$4.41 & 0.55$\pm$0.14 & 0 & 7.38$\pm$2.08 & 60.36$\pm$5.06 & 0.02$\pm$0.01 \\ \hline
   DDO161, B & 56.11$\pm$4.38 & 10.94$\pm$1.35 & 0.48$\pm$0.11 & 0 & 6.78$\pm$1.74 & 65.48$\pm$10.06 & 0.06$\pm$0.05 \\ \hline
   ESO079-G014, B & 118.04$\pm$4.46 & 20.57$\pm$1.22 & 0.47$\pm$0.11 & 0 & 22.33$\pm$6.37 & 78.1$\pm$4.92 & 0.18$\pm$0.21 \\ \hline
   ESO116-G012, B & 68.92$\pm$2.28 & 25.51$\pm$2.1 & 0.5$\pm$0.12 & 0 & 11.76$\pm$3.44 & 73.78$\pm$3.02 & 0.14$\pm$0.08 \\ \hline
   F583-4, B & 105.87$\pm$252.16 & 9.48$\pm$7.1 & 0.63$\pm$0.17 & 0 & 60.29$\pm$10.41 & 59.32$\pm$9.4 & 0.13$\pm$0.1 \\ \hline
   NGC 1090, B & 99.23$\pm$2.85 & 29.9$\pm$2.45 & 0.38$\pm$0.1 & 0 & 19.68$\pm$8.95 & 63.1$\pm$3.02 & 0.29$\pm$0.29 \\ \hline
   NGC 2403, B & 83.49$\pm$2.62 & 21.53$\pm$0.67 & 0.8$\pm$0.07 & 0 & 2.93$\pm$0.15 & 57.86$\pm$2.96 & 0.44$\pm$0.07 \\
\hline
   \end{longtable}
\newpage
\renewcommand\thetable{2}
\begin{longtable}{|| c || c | c | c | c | c | c | c |} 
 \centering
    Galaxy & $V_{200}$, km/s & $C_{200}$ & $Y_d$ & $Y_b$ & D, Mpc & i, $^o$ & $G\mu, 10^{-8}$ \\ \hline\hline
   UGC 03205, B & 216.1$\pm$171.82 & 4.49$\pm$1.42 & 00.65$\pm$0.1 & 0.48$\pm$0.11 & 51.14$\pm$8.4 & 66.67$\pm$4.15 & 11.63$\pm$2.04 \\ \hline
   UGC 04278, B & 120.15$\pm$120.7 & 14.43$\pm$1.46 & 0.63$\pm$0.16 & 0 & 9.93$\pm$2.16 & 88.01$\pm$1.4 & 0.09$\pm$0.07 \\ \hline 
   UGC 05005, B & 104.66$\pm$47.21 & 12.41$\pm$7.29 & 0.53$\pm$0.14 & 0 & 54.25$\pm$10.76 & 31.32$\pm$12.5 & 0.26$\pm$0.58 \\ \hline
   UGC 05716, NFW & 59.55$\pm$9.01 & 8.61$\pm$1.13 & 0.47$\pm$0.11 & 0 & 22.58$\pm$4.97 & 52.94$\pm$9.09 & 0.1$\pm$0.13 \\ \hline
   UGC 05716, B & 46.27$\pm$6.14 & 15.01$\pm$1.74 & 0.52$\pm$0.13 & 0 & 22.86$\pm$5.13 & 53.4$\pm$8.9 & 0.6$\pm$0.27 \\ \hline
   UGC 05721, B & 40.09$\pm$2.1 & 53.2$\pm$5.29 & 0.51$\pm$0.12 & 0 & 6.24$\pm$1.66 & 60.09$\pm$5.06 & 0.09$\pm$0.1 \\ \hline
   UGC 06614, B & 213.26$\pm$57.86 & 10.62$\pm$3.49 & 0.52$\pm$0.13 & 0.71$\pm$0.18 & 88.13$\pm$8.73 & 24.08$\pm$6.04 & 13.45$\pm$20.04 \\
\end{longtable}
\newpage
\renewcommand\thetable{3}
\begin{longtable}{|| c || c | c | c | c | c | c | c |} 
 \centering
    Galaxy & $V_{200}$, km/s & $C_{200}$ & $Y_d$ & $Y_b$ & D, Mpc & i, $^o$ & $G\mu, 10^{-8}$  \\ \hline\hline
   ESO116-G012 & 68.9$\pm$2.28 & 25.53$\pm$2.12 & 0.5$\pm$0.12 & 0 & 11.77$\pm$3.47 & 73.65$\pm$3 & 0.14$\pm$0.08 \\ \hline
   NGC 2403, B & 83.83$\pm$2.69 & 21.59$\pm$0.67 & 0.81$\pm$0.07 & 0 & 2.93$\pm$0.16 & 57.54$\pm$3.03 & 0.44$\pm$0.07 \\ \hline
   NGC 2841, NFW & 389.84$\pm$73.89 & 0.69$\pm$0.51 & 0.86$\pm$0.12 & 0.49$\pm$0.07 & 13.98$\pm$1.3 & 74.29$\pm$8.31 & 23.46$\pm$3.98 \\ \hline
   NGC 2841, B & 400.86$\pm$68.93 & 3.24$\pm$0.4 & 0.79$\pm$0.11 & 0.39$\pm$0.06 & 13.37$\pm$1.29 & 71.69$\pm$9.35 & 31.83$\pm$4.67 \\ \hline
   NGC 2903, B & 93.06$\pm$1.98 & 79.08$\pm$2.33 & 0.49$\pm$0.13 & 0 & 0.1$\pm$0.16 & 65.61$\pm$3.01 & 1.8$\pm$0.53 \\ \hline
   NGC 2955, NFW & 171.13$\pm$23.17 & 13$\pm$5.59 & 0.41$\pm$0.09 & 0.7$\pm$0.13 & 92.17$\pm$9.94 & 44.02$\pm$5.74 & 9.89$\pm$3.15 \\ \hline
   NGC 2955, B & 143.66$\pm$14.47 & 19.2$\pm$4 & 0.42$\pm$0.09 & 0.76$\pm$0.15 & 93.33$\pm$9.31 & 47.5$\pm$5.67 & 8.72$\pm$2.68 \\
\end{longtable}
\newpage
\renewcommand\thetable{4}
\begin{longtable}{|| c || c | c | c || c | c | c |} 
 \centering
    Galaxy & NFW, w/o string & NFW, string & NFW, finite string & B, w/o string & B, string & B, finite string  \\ \hline\hline
    NGC 2841 & 105.02 & 91.36 & 98.41 & 121.81 & 93.17 & 93.88 \\ \hline
    NGC 2955 & 98.16 & 73.1 & 72.83 & 91.85 & 68.87 & 68.63 \\ \hline
    NGC 2998 & 42.58 & 23.86 & 23.75 & 44.65 & 23.7 & 23.35 \\ \hline
    NGC 5371 & 88.94 & 40.78 & 42.72 & 161.36 & 40.63 & 38.9 \\ \hline
    NGC 5585 & 145.45 & 143.37 & 143.35 & 110.84 & 88.45 & 88.43 \\ \hline
    NGC 5907 & 95.8 & 63.1 & 65.16 & 107.7 & 59.96 & 59.49 \\ \hline
    UGC 06786 & 52.85 & 51.16 & 51.73 & 113.81 & 58.8 & 61.33 \\ \hline
    UGC 09133 & 477.21 & 439.18 & 440.31 & 531.1 & 428.28 & 434.16 \\ \hline
    UGC 11820 & 23.6 & 22.18 & 22.7 & 48.59 & 34.7 & 34.49 \\ \hline
    \caption{The values of the Bayesian criterion for three cases: spherical DM halo, spherical DM halo with an inifinte filament at the origin, and spherical DM halo with a finite filament of length 200 kpc at the origin. The values are given for 9 galaxies that demonstrate BIC improvement upon the addition of the filament for \textit{both} NFW and Burkert profiles (we refer to this as \textit{strong} evidence for the filament): the left portion of the table concerns NFW halos, while the right portion contains corresponding results for the Burkert halos.}
    \label{tab:table4}
\end{longtable}
\newpage
\renewcommand\thetable{5}
\begin{longtable}{|| c || c | c | c || c | c | c |} 
 \centering
    Galaxy & NFW, w/o string & NFW, string & NFW, finite string & B, w/o string & B, string & B, finite string  \\ \hline\hline
    ESO116-G012 & 44.01 & --- & --- & 23.54 & 22.88 & 22.85 \\ \hline
    NGC 2903 & 204.52 & --- & --- & 227.75 & 192.35 & 194.25 \\ \hline
    NGC 3198 & 74.66 & --- & --- & 76.75 & 68.93 & 70.32 \\ \hline
    NGC 6674 & 45.12 & --- & --- & 40.38 & 40.29 & 40.86 \\ \hline
    NGC 6946 & 114.4 & --- & --- & 111.69 & 103.48 & 103.09 \\ \hline
    UGC 05716 & 32.39 & 34.03 & --- & 40.36 & 27.73 & 28.42 \\ \hline
    \caption{The values of the Bayesian criterion for three cases: spherical DM halo, spherical DM halo with an inifinte filament at the origin, and spherical DM halo with a finite filament of length 200 kpc at the origin. The values are given for 6 galaxies that demonstrate BIC improvement for either NFW or Burkert profile, compared to the standard case for the \textit{better}-fitting profile (we refer to this as \textit{moderate} evidence for the filament): the left portion of the table concerns NFW halos, while the right portion contains corresponding results for the Burkert halos. The cases when the initial fit minimum is at $G\mu=0$ are marked with dashes.}
    \label{tab:table5}
\end{longtable}
\newpage
\renewcommand\thetable{6}
\begin{longtable}{|| c || c | c | c || c | c | c |} 
 \centering
    Galaxy & NFW, w/o string & NFW, string & NFW, finite string & B, w/o string & B, string & B, finite string  \\ \hline\hline
    NGC 2403 & 648.36 & --- & --- & 822.35 & 756.14 & 757.05  \\ \hline
    NGC 4183 & 19.34 & --- & --- & 23.72 & 23.17 & 23.24 \\ \hline
    NGC 5985 & 81.61 & --- & --- & 109.89 & 82.55 & 85.52 \\ \hline
    NGC 6503 & 58.48 & 61.26 & --- & 83.28 & 70.67 & 71.77 \\ \hline
    UGC 00128 & 76.63 & 86 & --- & 176.42 & 86 & 89.43 \\ \hline
    UGC 00731 & 15.81 & --- & --- & 17.18 & 16.47 & 16.45 \\ \hline
    UGC 03205 & 176.43 & 176.59 & --- & 190.57 & 184.33 & 185.54 \\ \hline
    UGC 08490 & 20.3 & --- & --- & 33 & 24.63 & 24.65 \\ \hline
    UGC 12732 & 17.56 & --- & --- & 24.16 & 22.52 & 22.57 \\ \hline
    \caption{The values of the Bayesian criterion for three cases: spherical DM halo, spherical DM halo with an inifinte filament at the origin, and spherical DM halo with a finite filament of length 200 kpc at the origin. The values are given for 9 galaxies that demonstrate BIC improvement for either NFW or Burkert profile, compared to the standard case for the \textit{worse}--fitting profile (we refer to this as \textit{weak} evidence for the filament): the left portion of the table concerns NFW halos, while the right portion contains corresponding results for the Burkert halos. The cases when the initial fit minimum is at $G\mu=0$ are marked with dashes.}
    \label{tab:table6}
\end{longtable}
\newpage
\renewcommand\thetable{7}
\begin{longtable}{|| c || c | c | c || c | c | c |} 
 \centering
    Galaxy & NFW, w/o string & NFW, string & B, w/o string & B, string  \\ \hline\hline
    DDO 064 & 19.67 & --- & 17.96 & 20.39 \\ \hline
    DDO 161 & 53.64 & --- & 27.23 & 29.37 \\ \hline
    ESO079-G014 & 61.49 & --- & 26.2 & 29.04 \\ \hline
    F583-4 & 22.18 & --- & 21.65 & 22.81 \\ \hline
    NGC 1090 & 68.71 & --- & 47.26 & 50.05 \\ \hline
    NGC 3109 & 210.82 & ---  & 20.31 & 23.01  \\ \hline
    NGC 3521 & 27.85 & 28 & 26.28 & 26.85 \\
\end{longtable}
\newpage
\renewcommand\thetable{8}
\begin{longtable}{|| c || c | c | c || c | c | c |} 
 \centering
    Galaxy & NFW, w/o string & NFW, string & NFW, finite string & B, w/o string & B, string & B, finite string  \\ \hline\hline
    NGC 2841 & 93.54 & 77.98 & 85.03 & 110.34 & 79.79 & 80.79 \\ \hline
    NGC 2955 & 91.1 & 64.86 & 64.58 & 84.78 & 60.62 & 60.39 \\ \hline
    NGC 2998 & 39.76 & 20.47 & 20.36 & 41.82 & 23.7 & 19.96 \\ \hline
    NGC 3521 & 19.28 & 17.7 & 17.66 & 17.71 & 16.57 & 16.59 \\ \hline
    NGC 5371 & 84.22 & 35.11 & 37.06 & 156.64 & 34.97 & 33.23 \\ \hline
    NGC 5585 & 139.56 & 136.3 & 136.28 & 104.95 & 81.38 & 81.36 \\ \hline
    NGC 5907 & 91.08 & 57.43 & 59.49 & 102.98 & 54.29 & 53.82 \\ \hline
    UGC 03205 & 165.2 & 163.49 & 164.42 & 179.34 & 171.23 & 172.44 \\ \hline
    UGC 06786 & 42.01 & 38.51 & 39.08 & 102.97 & 46.16 & 48.69 \\ \hline
    UGC 09133 & 463.9 & 423.65 & 424.77 & 517.78 & 412.74 & 418.62 \\ \hline
    UGC 11820 & 22.1 & 20.37 & 20.88 & 47.08 & 32.88 & 32.67 \\ \hline
    \caption{The values of the Akaike criterion for three cases: spherical DM halo, spherical DM halo with an infinite filament at the origin, and spherical DM halo with a finite filament of length 200 kpc at the origin. The values are given for 11 galaxies that demonstrate AIC improvement upon the addition of the filament for \textit{both} NFW and Burkert profiles (we refer to this as \textit{strong} evidence for the filament): the left portion of the table concerns NFW halos, while the right portion contains corresponding results for the Burkert halos.}
    \label{tab:table8}
\end{longtable}
\newpage
\renewcommand\thetable{9}
\begin{longtable}{|| c || c | c | c || c | c | c |} 
 \centering
    Galaxy & NFW, w/o string & NFW, string & NFW, finite string & B, w/o string & B, string & B, finite string  \\ \hline\hline
    ESO116-G012 & 40.47 & --- & --- & 20 & 18.64 & 18.6 \\ \hline
    NGC 2903 & 196.89 & --- & --- & 220.12 & 183.19 & 185.09 \\ \hline
    NGC 3198 & 65.85 & --- & --- & 67.95 & 58.37 & 59.76 \\ \hline
    NGC 5985 & 72.63 & --- & --- & 100.91 & 72.08 & 75.04 \\ \hline
    NGC 6674 & 40.87 & --- & --- & 36.13 & 35.33 & 35.91 \\ \hline
    NGC 6946 & 102.04 & --- & --- & 99.33 & 89.05 & 88.67 \\ \hline
    UGC 05716 & 29.96 & 31.12 & --- & 37.93 & 24.83 & 25.51 \\ \hline
    \caption{The values of the Akaike criterion for three cases: spherical DM halo, spherical DM halo with an inifinte filament at the origin, and spherical DM halo with a finite filament of length 200 kpc at the origin. The values are given for 7 galaxies that demonstrate AIC improvement for either NFW or Burkert profile, compared to the standard case for the \textit{better}-fitting profile (we refer to this as \textit{moderate} evidence for the filament): the left portion of the table concerns NFW halos, while the right portion contains corresponding results for the Burkert halos. The cases when the initial fit minimum is at $G\mu=0$ are marked with dashes.}
    \label{tab:table9}
\end{longtable}
\newpage
\renewcommand\thetable{10}
\begin{longtable}{|| c || c | c | c || c | c | c |} 
 \centering
    Galaxy & NFW, w/o string & NFW, string & NFW, finite string & B, w/o string & B, string & B, finite string  \\ \hline\hline
    NGC 2403 & 636.91 & --- & --- & 810.89 & 742.39 & 743.3 \\ \hline
    NGC 4183 & 13.66 & --- & --- & 18.04 & 16.36 & 16.42 \\ \hline
    NGC 6503 & 51.31 & 52.65 & --- & 76.11 & 62.06 & 63.16 \\ \hline
    UGC 00128 & 71.18 & --- & --- & 170.96 & 79.45 & 82.88 \\ \hline
    UGC 00731 & 13.38 & --- & --- & 14.76 & 13.56 & 13.54 \\ \hline
    UGC 08490 & 13.3 & --- & --- & 26 & 16.22 & 16.24 \\ \hline
    UGC 12732 & 13.7 & --- & --- & 20.3 & 17.89 & 17.93 \\ \hline
    \caption{The values of the Akaike criterion for three cases: spherical DM halo, spherical DM halo with an inifinte filament at the origin, and spherical DM halo with a finite filament of length 200 kpc at the origin. The values are given for 7 galaxies that demonstrate AIC improvement for either NFW or Burkert profile, compared to the standard case for the \textit{worse}--fitting profile (we refer to this as \textit{weak} evidence for the filament): the left portion of the table concerns NFW halos, while the right portion contains corresponding results for the Burkert halos. The cases when the initial fit minimum is at $G\mu=0$ are marked with dashes.}
    \label{tab:table10}
\end{longtable}
\newpage
\renewcommand\thetable{11}
\begin{longtable}{|| c || c | c | c || c | c | c |} 
 \centering
    Galaxy & NFW, w/o string & NFW, string & B, w/o string & B, string \\ \hline\hline
    DDO064 & 16.47 & --- & 14.76 & 16.55 \\ \hline
    DDO 161 & 46.47 & --- & 20.06 & 20.77 \\ \hline
    ESO079-G014 & 57.95 & --- & 22.66 & 24.79 \\ \hline
    F583-4 & 19.76 & --- & 19.23 & 19.9 \\ \hline
    NGC 1090 & 62.82 & --- & 41.37 & 43 \\ \hline
    NGC 3109 & 204.72 & --- & 14.21 & 15.7 \\ \hline
    NGC 3972 & 19.31 & --- & 16.04 & 18.72 \\
\end{longtable}
\newpage
\renewcommand\thetable{12}
\begin{longtable}{|| c || c | c | c | c | c | c | c | c | c | c | c | c | c | c |} 
 \centering
 \small
    Galaxy & $V_{200}$, km/s & $C_{200}$ & $Y_d$ & $Y_b$ & D, Mpc & i, $^o$ & q & BIC & AIC \\ \hline\hline
   NGC 5371, NFW & 131.73$\pm$5.21 & 26.9$\pm$1.67 & 0.38$\pm$0.09 & 0 & 27.4$\pm$6.76 & 52.46$\pm$2.03 & 0.41$\pm$0.18 & 86.81 & 81.14 \\ \hline
   NGC 5371, B & 440.08$\pm$43.49 & 3.25$\pm$0.22 & 0.86$\pm$0.14 & 0 & 30.96$\pm$5.02 & 52.76$\pm$1.97 & 0.7$\pm$0.21 & 158.5 & 152.83 \\ \hline
   NGC 5907, NFW & 144.76$\pm$2.94 & 23.21$\pm$2.05 & 0.28$\pm$0.05 & 0 & 16.6$\pm$0.89 & 87.59$\pm$1.5 & 0.44$\pm$0.21 & 94.36 & 88.69 \\ \hline
   NGC 5907, B & 202.33$\pm$23.55 & 5.6$\pm$0.49 & 0.96$\pm$0.05 & 0 & 17.82$\pm$0.86 & 87.62$\pm$1.47 & 0.66$\pm$0.23 & 110.77 & 105.1 \\ \hline
    \caption{The values of fitting parameters with errors, BIC and AIC for the 2 galaxies, obtained from the MCMC analysis, for a deformed dark matter profile.}
\label{tab:table12}
\end{longtable}
\newpage
\renewcommand\thetable{12}
\begin{longtable}{|| c | c | c | c | c | c | c | c | c | c | c | c |} 
 \centering
    $V_{200}$, km/s & $C_{200}$ & $Y_d$ & $Y_b$ & D, Mpc & i, $^o$ & $G\mu, 10^{-8}$ & q & BIC & AIC \\ \hline\hline
    307.71$\pm$141.33 & 1.46$\pm$0.64 & 0.5$\pm$0.06 & 0 & 17.05$\pm$0.86 & 87.5$\pm$1.52 & 15.84$\pm$1.48 & 0.68$\pm$0.22 & 62.59 & 55.98 \\ \hline
    \caption{The values of fitting parameters with errors, BIC and AIC for NGC 5907 (Burkert profile), obtained from the MCMC analysis, for a deformed dark matter profile plus a filament.}
\label{tab:table13}
\end{longtable}
\end{document}